\documentclass[conference]{IEEEtran}
\IEEEoverridecommandlockouts
\usepackage{cite}
\usepackage{amsmath,amssymb,amsfonts}
\usepackage{graphicx}
\usepackage{textcomp}
\usepackage{booktabs}
\usepackage{tabularx}
\usepackage{tikz}
\usetikzlibrary{positioning,arrows.meta,backgrounds,calc}
\usepackage{xcolor}

\def\BibTeX{{\rm B\kern-.05em{\sc i\kern-.025em b}\kern-.08em T\kern-.1667em\lower.7ex\hbox{E}\kern-.125emX}}

\begin{document}

\title{Experimental Classification of xApp Conflicts in O-RAN: Slicing vs. Per-UE Scheduling}

\author{%
\IEEEauthorblockN{
Rashmi Sharan Sinha\IEEEauthorrefmark{1},
Merim Dzaferagic\IEEEauthorrefmark{2}, Marco Ruffini\IEEEauthorrefmark{2}, and
Nima Afraz\IEEEauthorrefmark{1}}
\IEEEauthorblockA{
\IEEEauthorrefmark{1}School of Computer Science,
University College Dublin, Dublin, Ireland
}
\IEEEauthorblockA{
\IEEEauthorrefmark{2}School of Computer Science and Statistics,
Trinity College Dublin, Dublin, Ireland
}
}
\maketitle

\begin{abstract}
Concurrent applications (xApps) in the Near-Real-Time RAN Intelligent Controller (Near-RT RIC) of the Open Radio Access Network (O-RAN) can issue control intents that are jointly infeasible when they act on hierarchically coupled radio resources. This paper experimentally studies an interaction between slice-level physical resource block (PRB) allocation and per-user-equipment (UE) scheduling. We implement a utilisation-aware slicing control (the spectral-efficiency xApp) that reclaims a fixed fraction of the baseline allocation from under-utilised slices and redistributes pooled capacity to saturated slices. The control logic is exercised in simulation and on a 5G testbed comprising an OCUDU-based next-generation NodeB (gNB), an Open5GS core, a Universal Software Radio Peripheral (USRP) X310, and five commercial off-the-shelf (COTS) UEs. Because the gNB counters available in this testbed are UE-keyed but not slice-labelled, the slicing control derives a per-slice throughput proxy from core-side traffic measurements. A command-bus detector evaluates explicitly defined control-intent conditions: a per-UE floor above its parent-slice cap, aggregate UE floors above the cell budget, and repeated slice-target changes within a specified window. Over a 276-cycle continuous run, the detector logged 147 per-slice floor--cap violations, 49 shared-budget over-subscription episodes, and 11 allocation-oscillation episodes. Against the conventional static slice split replayed on the same load trace, the slicing control reduces unmet demand by 12--33\,\% and idle allocation by 21--23\,\%. The results demonstrate reproducible cross-granularity control-intent interactions on real hardware and provide an experimental basis for analysing conflict detection in O-RAN. 

\end{abstract}


\section{Introduction}
The Open Radio Access Network (O-RAN) architecture enables independently developed applications (xApps) to monitor and control RAN functions through the Near-Real-Time RAN Intelligent Controller (Near-RT RIC) using standardised interfaces~\cite{oran_arch,polese_survey}. This programmability supports flexible, multi-vendor control and forms part of the broader evolution towards artificial-intelligence (AI)-powered network control, in which autonomous decisions must eventually be coordinated across multiple network domains, functions, and control timescales~\cite{dzaferagic2025decentralized}. However, this programmability also creates a feasibility problem when xApps issue decisions over shared radio resources at different control granularities. In particular, a slice-level xApp can configure the physical resource block (PRB) budget of a network slice while a per-user-equipment (UE) scheduling xApp assigns a PRB guarantee to a UE within that slice. Because each UE-level allocation must remain feasible within the budget of its parent slice, a UE floor above the corresponding slice cap produces an incompatible control state. Concurrent UE floors can also exceed the shared cell-wide PRB budget.


Experimental studies have demonstrated xApp conflicts on physical O-RAN testbeds where applications act at the same resource-management level, but cross-granularity interactions between slice-level allocation and per-UE scheduling remain insufficiently characterised experimentally.

In this paper, we implement and evaluate cross-granularity interaction
between slice-level reallocation and per-UE scheduling, first in
simulation and then on a physical O-RAN testbed. A command-bus detector
evaluates the resulting control-intent conditions, while a core-derived
throughput proxy provides per-slice load information when slice-labelled measurements from the next-generation NodeB (gNB) are unavailable.


Our contributions are as follows.
\begin{itemize}
\item We implement and evaluate a cross-granularity O-RAN control scenario that combines slice-level reallocation with per-UE scheduling in simulation and on a physical 5G testbed.


\item We define and evaluate a command-bus detector for parent-slice floor--cap violations, shared-budget over-subscription, and allocation-oscillation events, and show that the budget-conserving slice policy cannot create shared-budget over-subscription in isolation.
\end{itemize}

\begin{figure}[t]
\centering
\resizebox{0.62\columnwidth}{!}{%
\begin{tikzpicture}[
    >=Latex,
    font=\small,
    xapp/.style={
        draw,
        rounded corners,
        fill=white,
        align=center,
        minimum width=2.7cm,
        minimum height=1.05cm
    },
    target/.style={
        draw,
        fill=white,
        align=center,
        minimum width=2.3cm,
        minimum height=0.95cm
    },
    system/.style={
        draw,
        fill=white,
        align=center,
        minimum height=0.9cm
    },
    bus/.style={
        draw,
        rounded corners,
        fill=white,
        align=center,
        minimum height=0.9cm
    },
    ctrl/.style={->, thick},
    feedback/.style={->, dashed, thick},
    logging/.style={->, dotted, thick}
]

\node[xapp] (a) at (0,4.5)
{\textbf{xApp A}\\Equal Slicing\\{\scriptsize baseline setup}};

\node[xapp] (b) at (3.35,4.5)
{\textbf{xApp B}\\Spectral Efficiency\\{\scriptsize dynamic slice control}};

\node[xapp] (c) at (6.7,4.5)
{\textbf{xApp C}\\UE Scheduler\\{\scriptsize per-UE guarantee}};

\node[target] (ta) at (0,2.95)
{Slice-level\\PRB baseline};

\node[target] (tb) at (3.35,2.95)
{Slice-level\\PRB target};

\node[target] (tc) at (6.7,2.95)
{Per-UE\\PRB floor};

\draw[ctrl] (a.south) -- node[right,font=\scriptsize]{initial} (ta.north);
\draw[ctrl] (b.south) -- node[right,font=\scriptsize]{dynamic} (tb.north);
\draw[ctrl] (c.south) -- node[right,font=\scriptsize]{dynamic} (tc.north);

\draw[feedback] ([yshift=4pt]a.east) -- ([yshift=4pt]b.west);

\node[system, minimum width=4.1cm] (gnb) at (3.35,1.25)
{\textbf{5G gNB}\\
{\scriptsize OCUDU + USRP X310}};

\draw[ctrl] (ta.south) |- 
    node[pos=.72,above,font=\scriptsize]{slice policy}
    (gnb.west);

\draw[ctrl] (tb.south) --
    node[right,font=\scriptsize]{slice policy}
    (gnb.north);

\draw[ctrl] (tc.south) |-
    node[pos=.72,above,font=\scriptsize]{UE policy}
    (gnb.east);

\node[system, minimum width=2.3cm] (core) at (0.65,-0.05)
{\textbf{5G Core}\\Open5GS};

\node[system, minimum width=2.3cm] (ue) at (6.05,-0.05)
{\textbf{COTS UEs}\\3 slices};

\draw[<->,thick] (core) -- node[above,font=\scriptsize]{N2/N3} (gnb.west);
\draw[<->,thick] (gnb.east) -- node[above,font=\scriptsize]{RF} (ue);

\node[bus, minimum width=7.0cm] (bus) at (3.35,-1.65)
{\textbf{Command bus}\hspace{3mm}{\scriptsize logged xApp control intents}};

\node[system, minimum width=3.2cm] (det) at (3.35,-3.0)
{\textbf{Conflict Detector}\\{\scriptsize classify control intents}};

\draw[ctrl] (bus.south) -- (det.north);

\begin{scope}[on background layer]

\draw[feedback]
    (core.west) -- ++(-0.55,0) 
    |- node[pos=0.25,left,font=\scriptsize,align=center]{per-slice\\load}
       ([xshift=-3pt]b.west);

\draw[logging]
    ([xshift=-6pt]a.west) -- ++(-0.55,0) |- (bus.west);

\draw[logging]
    (b.south) -- (bus.north);

\draw[logging]
    ([xshift=6pt]c.east) -- ++(0.55,0) |- (bus.east);

\end{scope}

\end{tikzpicture}%
}
\caption{Experimental architecture.}
\label{fig:arch}
\end{figure}
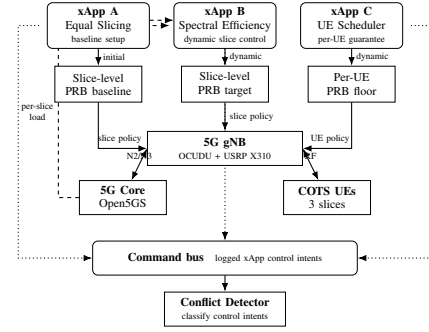


\section{Related Work}\label{sec:related}

Platforms such as SCOPE~\cite{scope} and OpenRAN Gym~\cite{openrangym}, with open-source stacks such as srsRAN~\cite{srsran} and Open5GS~\cite{open5gs}, have enabled xApp evaluation on software-defined radios~\cite{usrp}. Conflict management among xApps has received increasing attention: Wadud \emph{et al.} formulated xApp coordination in the Near-RT RIC as a game-theoretic problem~\cite{wadud2023conflict}, introduced quality-of-service (QoS)-aware mitigation through QACM~\cite{wadud2024qacm}, studied xApp-level mitigation for interacting mobility and energy-saving loops~\cite{wadud2024xapp}, and considered AI-assisted detection, classification, and mitigation~\cite{wadud2026ai}. These works address mechanisms for resolving conflicting objectives in simulation or on analytical examples.

Closest to this work, Sultana \emph{et al.} demonstrated direct-conflict detection and mitigation with a single UE on an end-to-end testbed with an O-RAN SC Near-RT RIC, srsRAN, Open5GS, and a USRP X310, comparing xApp control messages that address the same target~\cite{10773739}, and later evaluated detection and mitigation over the air with three UEs and two xApps independently modifying slice-level PRB allocations~\cite{11152996}. Their interactions concern xApps acting on the same slice-level target.

Table~\ref{tab:related} positions the present work. It addresses a different gap: xApps operating at different but hierarchically coupled granularities, where a slice-level control sets the PRB budget of a slice and a per-UE scheduler requests a floor for a UE within it, so that the control intents must satisfy both a parent-slice feasibility condition and the shared cell budget. We reproduce these cross-granularity conditions on hardware with a core-derived per-slice load proxy that survives UE radio-identifier changes, and distinguish the budget-conserving slice policy from the over-subscription introduced by concurrent per-UE floors.

\begin{table}[t]
\caption{Comparison with prior work on xApp conflicts. SDR: software-defined radio; OTA: over the air; MRO: mobility robustness optimisation; QoS: quality of service.}
\label{tab:related}
\centering
\scriptsize
\setlength{\tabcolsep}{2.5pt}
\renewcommand{\arraystretch}{1.1}
\begin{tabularx}{\columnwidth}{@{}l>{\raggedright\arraybackslash}p{1.35cm}>{\raggedright\arraybackslash}p{1.45cm}>{\raggedright\arraybackslash}p{1.35cm}>{\raggedright\arraybackslash}X@{}}
\toprule
Work & Platform & Interacting xApps & Conflict classes & Scope \\
\midrule
\cite{wadud2023conflict} & Analytical example & Same parameter, same level & Direct, indirect & Detection, mitigation (game-theoretic) \\
\cite{wadud2024qacm} & Simulation & Same parameter, same level & Direct, indirect, implicit & QoS-aware mitigation \\
\cite{wadud2024xapp} & Simulation & MRO vs.\ energy saving & Direct & Detection, mitigation \\
\cite{wadud2026ai} & Simulation & 5--50 synthetic xApps & Direct, indirect, implicit & AI detection, classification \\
\cite{10773739} & Testbed, SDR, 1 COTS UE & Two, same control target & Direct & Detection, mitigation (demo) \\
\cite{11152996} & Testbed, OTA, 3 COTS UEs & Two, slice-level PRB & Direct & Detection, mitigation \\
Our work & Testbed, OTA, 5 COTS UEs & Slice-level and per-UE (cross-granularity) & Direct, indirect, implicit & Detection; core-derived load proxy; static-split comparison \\
\bottomrule
\end{tabularx}
\vspace{-6pt}
\end{table}

\section{System Model and Background}\label{sec:model}

We consider a single 5G cell serving a set of network slices
$\mathcal{S}=\{1,2,3\}$, identified by Slice/Service Type (SST) 1, 2, and 3. The total downlink
PRB budget of the cell is normalised to $100\,\%$. Let $c_s$ denote the
PRB share assigned to slice $s\in\mathcal{S}$, and let $\mathcal{U}_s$
denote the set of UEs associated with that slice.

At the \emph{slice level}, the slicing xApp configures the minimum and maximum PRB shares of each slice; here $\texttt{min}_s=\texttt{max}_s=c_s$, so the configured value acts as a fixed allocation during a control cycle, subject to
\begin{equation}
    \sum_{s\in\mathcal{S}} c_s \leq 100.
    \label{eq:slice_budget}
\end{equation}

At the \emph{UE level}, the scheduling xApp assigns a guaranteed PRB floor $g_u$ to UE $u$, which belongs to a parent slice $s(u)\in\mathcal{S}$ and is addressed by its gNB distributed-unit F1 Application Protocol (gNB-DU-UE-F1AP) identifier. Since UE-level guarantees are realised within the parent-slice budget,
\begin{equation}
    \sum_{u\in\mathcal{U}_s} g_u \leq c_s,
    \qquad \forall s\in\mathcal{S}.
    \label{eq:hierarchical_feasibility}
\end{equation}
which, with one scheduled UE per slice as here, reduces to
\begin{equation}
    g_u \leq c_{s(u)}.
    \label{eq:ue_slice_feasibility}
\end{equation}

Interpreted as simultaneous guarantees, the floors are also coupled through the cell budget,
\begin{equation}
    \sum_{u\in\mathcal{U}} g_u \leq 100.
    \label{eq:global_feasibility}
\end{equation}

Violations of \eqref{eq:hierarchical_feasibility}--\eqref{eq:global_feasibility} indicate incompatible control intents between the slice-level and UE-level xApps. 

\section{Proposed xApps}\label{sec:xapps}
Fig.~\ref{fig:arch} summarises the three xApps and their corresponding
control targets.

\subsection{xApp~A: Equal Slicing Divider}

xApp~A initialises the slice-level PRB allocation as $\mathbf{B}=[34,33,33]$.
setting $\texttt{min}_s=\texttt{max}_s=B_s$ for each slice, and is then disabled, leaving xApp~B as the sole writer of the slice-level policy.\subsection{xApp~B: Spectral-Efficiency Reallocation}

xApp~B performs periodic slice-level PRB reallocation with a configured inter-cycle sleep of 2\,s. Because each cycle also contains a one-second byte-count sampling window and three policy pushes to the gNB, the achieved control period on the testbed is about 3.7\,s (median 4\,s); all durations reported in Section~\ref{sec:results} are measured from log timestamps, while cycle counts are exact. Each cycle it (i) estimates the utilisation of each slice, (ii) computes target shares by the policy below, reclaiming from under-utilised slices and feeding saturated ones, and (iii) applies them at the gNB.

\textit{Utilisation relative to the baseline.}
For slice~$i$ with baseline allocation $B_i$ and measured usage $u_i$,
both expressed as percentages of the total cell capacity, we define the
normalised utilisation as
\begin{equation}
    \rho_i = \frac{u_i}{B_i}.
    \label{eq:utilisation}
\end{equation}
The utilisation is referenced to the fixed baseline $B_i$ rather than the current reallocated share, so that the control decision itself does not change the reference in the next cycle; normalising against a recently reduced allocation would make an under-utilised slice appear immediately more loaded and induce unnecessary switching between allocation states.

\textit{Decision rules.}
The reallocation policy is summarised in Table~\ref{tab:rules}. A slice
with $\rho_i<0.5$ is classified as under-utilised, and $25\,\%$ of its
baseline allocation is reclaimed into a shared resource pool. For
$0.5\leq\rho_i<1$, the baseline allocation is retained. A slice with
$\rho_i\geq1$ is classified as saturated and is assigned an additional
five percentage points, provided that the shared pool contains at least
five percentage points of available capacity.

\begin{table}[t]
\caption{Decision rules of the spectral-efficiency xApp.}
\label{tab:rules}
\centering
\scriptsize
\setlength{\tabcolsep}{2pt}
\renewcommand{\arraystretch}{1.0}
\begin{tabularx}{\columnwidth}{@{}llX@{}}
\toprule
$\rho_i$ & State & Action on slice $i$ \\
\midrule
$\rho_i < 0.5$      & Under-utilised & Target $0.75B_i$; reclaim $0.25B_i$ to pool. \\
$0.5 \le \rho_i < 1$ & Nominal        & Retain baseline $B_i$. \\
$\rho_i \ge 1$       & Saturated      & Target $B_i{+}5$ if pool $\ge 5$ pts. \\
\bottomrule
\end{tabularx}
\vspace{-10pt}
\end{table}


\textit{Smoothing.} Because real traffic is bursty, the xApp averages the last three readings before applying the rules.

\textit{Budget conservation.} An important property of the rules is that the sum of the targets plus the pool always equals 100\,\%. Reclaiming moves PRBs from a slice into the pool without changing the total, and feeding moves PRBs from the pool to a slice, again without changing the total. Formally, if $T_i$ is the target of slice~$i$ and $P$ is the pool,
{\setlength{\abovedisplayskip}{2pt}\setlength{\belowdisplayskip}{2pt}%
\setlength{\abovedisplayshortskip}{2pt}\setlength{\belowdisplayshortskip}{2pt}
\begin{equation}
\sum_i T_i + P = \sum_i B_i = 100\,\%.
\label{eq:cons}
\end{equation}}

The xApp therefore never allocates more PRBs than the cell contains; Section~\ref{sec:conflict} shows that this property determines which conflicts the design can create.

The full procedure is listed in Algorithm~1.

\begin{figure}[t]
\centering
\fbox{\begin{minipage}{0.92\columnwidth}
\textbf{Algorithm 1}: Spectral-efficiency xApp (one cycle)\\[3pt]
\hrule
\vspace{4pt}
\footnotesize
1:\; read baseline $B_1,\dots,B_n$ from file\\
2:\; $u_i \gets$ smoothed measured usage of slice $i$\\
3:\; $T_i \gets B_i$ for all $i$;\quad $P \gets 0$\\
4:\; \textbf{for} each slice $i$ \textbf{do} \hfill\emph{(extract)}\\
5:\; \quad \textbf{if} $u_i/B_i < 0.5$ \textbf{then}\\
6:\; \quad\quad $T_i \gets 0.75\,B_i$;\quad $P \gets P + 0.25\,B_i$\\
7:\; \textbf{for} each slice $i$ with $u_i/B_i \ge 1.0$ \textbf{do} \hfill\emph{(feed)}\\
8:\; \quad \textbf{if} $P \ge 5$ \textbf{then} $T_i \gets T_i + 5$;\; $P \gets P - 5$\\
9:\; write $T_i$ to the gNB for every slice
\end{minipage}}
\label{alg:se}
\vspace{-10pt}
\end{figure}

\subsection{xApp~C: Per-UE Scheduling}

xApp~C implements UE-level resource control by assigning a guaranteed
PRB floor $g_u$ to a selected UE $u$. In the experimental configuration,
one instance of xApp~C is associated with each slice, and each instance
sets $g_u = 40\,\%$. The resulting UE-level constraint is evaluated
against the PRB cap $c_{s(u)}$ of the UE's parent slice, as defined in
Section~\ref{sec:model}. Thus, whenever $g_u > c_{s(u)}$, the UE-level
guarantee and slice-level allocation are not jointly feasible.

Since the focus is detection and classification rather than resolution, xApp~C applies its UE-level control independently of xApp~B, without any mitigation, so that the cross-granularity interactions can be observed and classified from the logged intents.

\section{Conflict Taxonomy and Detection}\label{sec:conflict}

Building on the established O-RAN conflict categories, we define the detector-observable conditions of Table~\ref{tab:conflicts_def}, tailored to the interaction between slice-level PRB allocation and per-UE scheduling, and identify conflict events from the logged xApp control intents.

\begin{table}[t]
\caption{Operational interpretation of the conflict classes used by the detector, and episodes detected over the 276-cycle continuous run (20\,s de-duplication).}
\label{tab:conflicts_def}
\centering
\footnotesize
\setlength{\tabcolsep}{3pt}
\renewcommand{\arraystretch}{1.12}
\begin{tabularx}{\columnwidth}{@{}lXlr@{}}
\toprule
Class & Interpretation & Condition & Count \\
\midrule
Direct
& A UE-level floor exceeds the cap of its parent slice & $g_u > c_{s(u)}$ & 147 \\

Indirect
& Aggregate UE guarantees exceed the shared cell budget & $\sum_u g_u > 100\,\%$ & 49 \\

Implicit
& Repeated changes in a slice target occur within the detector window & $\geq 4$ target changes & 11 \\

\bottomrule
\end{tabularx}
\end{table}
Each xApp records its control intent (timestamp, source xApp, control target, slice and UE identifiers, requested PRB value) on the shared command bus of Fig.~\ref{fig:arch}; commands within one observation interval are evaluated jointly against the conditions in Table~\ref{tab:conflicts_def}. 
A consequence of the conservation property~\eqref{eq:cons} is that the proposed spectral-efficiency xApp \emph{cannot} create an indirect conflict on its own. Because its per-slice targets always sum to at most 100\,\%, the shared budget is never over-subscribed, and the indirect condition is never met. To register an indirect conflict one needs an actuator that guarantees, across several slices, more than the whole budget---for example, the per-UE scheduler guaranteeing floors on three slices that sum beyond 100\,\%, together with a detector that counts per-UE floors toward the shared-budget check. We construct exactly this situation in Section~\ref{sec:results}.

\section{Experimental Setup}\label{sec:setup}
\begin{figure}[t]
\centering
\setlength{\abovecaptionskip}{3pt}   
\setlength{\belowcaptionskip}{0pt}
\includegraphics[width=0.72\columnwidth]{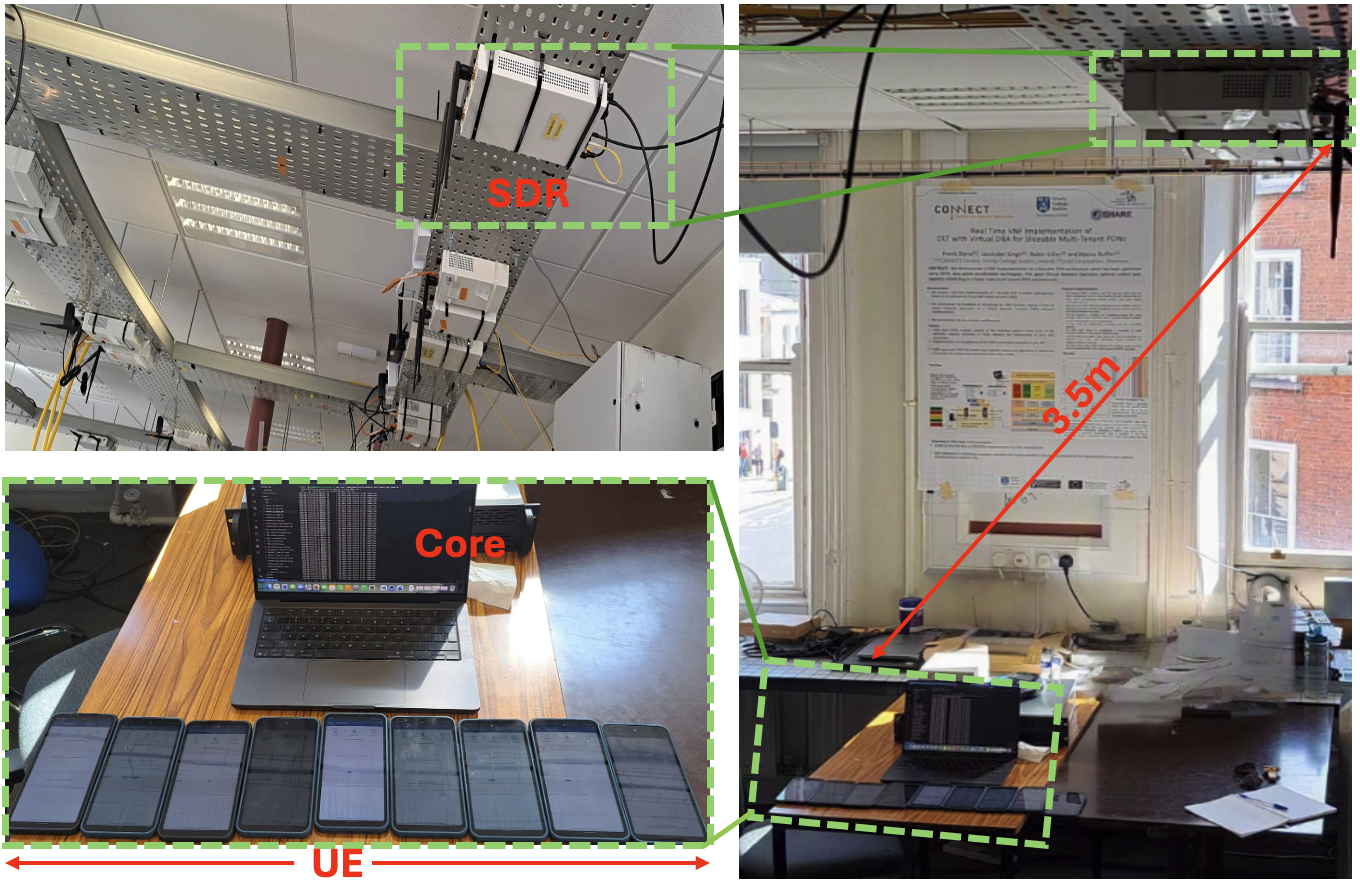}
\caption{Testbed setup.}
\vspace{-10pt}
\label{fig:setup}

\label{fig:testbed}
\end{figure}

The testbed comprises an Ettus Universal Software Radio Peripheral (USRP) X310 software-defined radio (SDR) driven by an OCUDU-based gNB (a centralised unit/distributed unit (CU/DU) running on a server; OCUDU is the open-source continuation of the srsRAN CU/DU), an Open5GS 5G standalone (SA) core, and commercial off-the-shelf (COTS) smartphones acting as UEs across the three slices. Fig.~\ref{fig:testbed} shows the physical setup: the radio is ceiling-mounted about 3.5\,m above the table holding the core/RIC laptop and the five handsets. The radio operates in band n78 with a 40\,MHz channel. xApps communicate with the gNB over its remote-control interface. Table~\ref{tab:testbed} lists the main parameters.

\begin{table}[t]
\caption{Testbed parameters.}
\label{tab:testbed}
\centering
\begin{tabular}{ll}
\toprule
Parameter & Value\\
\midrule
Radio & USRP X310, band n78, 40\,MHz\\
gNB & OCUDU-based CU/DU\\
Core & Open5GS 5G SA\\
User equipment & 5 COTS smartphones (UEs across 3 slices)\\
Slices & SST 1, 2, 3\\
Baseline split & 34 / 33 / 33\,\%\\
Reclaim / feed & 25\,\% of baseline / +5 points\\
Control cadence & 2\,s sleep ($\approx$3.7\,s achieved); smoothing over 3 cycles\\
\bottomrule
\end{tabular}
\vspace{-10pt}
\end{table}

\subsection{Core-Based Per-Slice Measurement}
The gNB exposes UE-level PRB counters, but they are not slice-labelled and are indexed by radio identifiers that change after UE re-attachment, so a persistent mapping to slices cannot be assumed. We therefore measure at the core, where a subscriber's slice and address are stable across radio re-attachments. Each phone's slice is read from the subscriber database (subscriber identity to SST) and its current address from the session-management logs (subscriber identity to IP), giving an address-to-slice map. Downlink bytes per address are counted at the core over a one-second window and summed per slice, yielding a per-slice throughput that serves as a proxy for PRB usage.

Concretely, let $r_i$ be the measured downlink rate of slice~$i$ over the sampling window and $C$ the cell's deliverable capacity. We express the slice's usage as a percentage of the whole cell,
\begin{equation}
u_i = 100\,\frac{r_i}{C},
\end{equation}
and feed $u_i$ into the utilisation $\rho_i = u_i/B_i$ used by the rules of Section~\ref{sec:xapps}. $C$ is set to the achievable rather than the nominal capacity, since the throughput of a software-defined radio is bounded well below the theoretical maximum.

This is a throughput proxy, not a PRB measurement, but it is immune to the radio-identifier churn that defeats the gNB-side counters. The map is rebuilt every 10\,s, so a phone that re-attaches with a new address is picked up automatically.

\section{Results and Discussion}\label{sec:results}
We evaluate the design first in simulation and then on the hardware testbed.

\subsection{Simulation}
\begin{figure}[t]
\centering
\includegraphics[width=0.88\columnwidth]{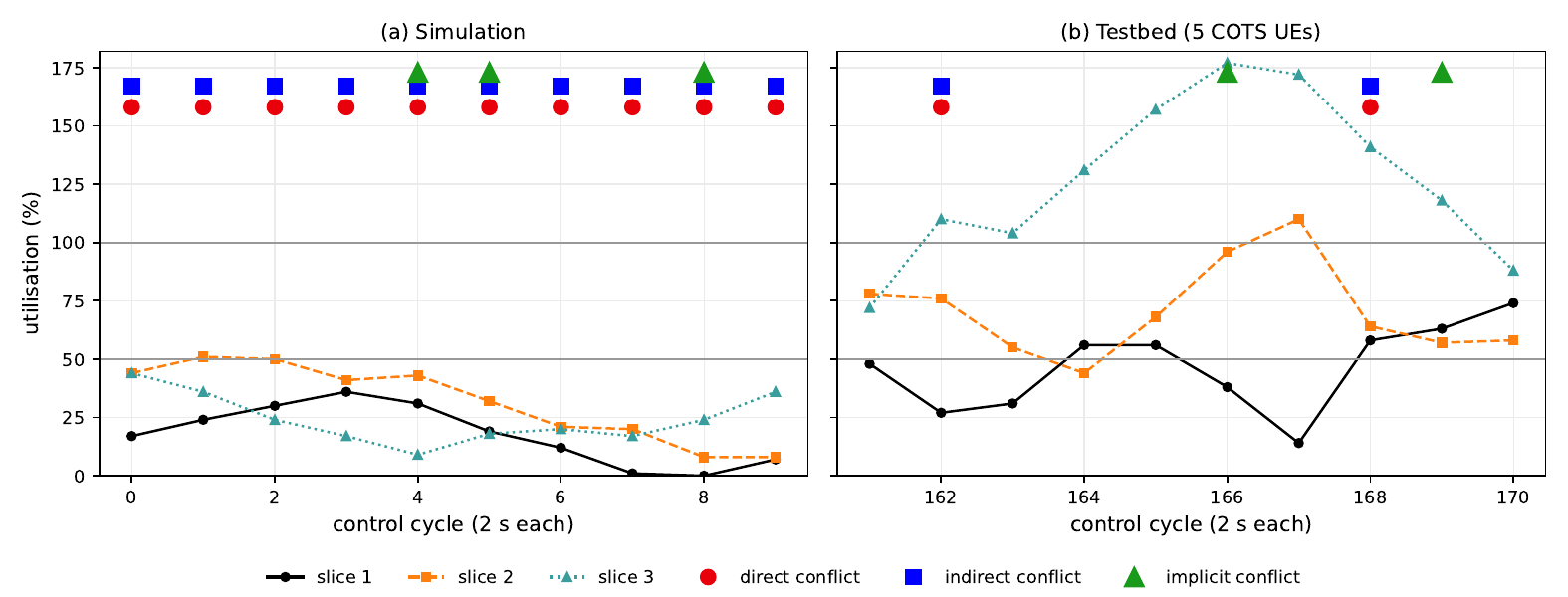}
\caption{Spectral-efficiency xApp over ten control cycles: (a) simulation, cycles 0--9, with synthetic load; (b) testbed, cycles 161--170 (33\,s), with five COTS UEs.}

\label{fig:simhw}
\end{figure}

Before deploying on hardware, we ran the spectral-efficiency xApp in simulation, driving the real decision logic of Algorithm~1 with a synthetic per-slice load model and the same three per-UE scheduler floors. Fig.~\ref{fig:simhw}(a) shows ten control cycles. In simulation the synthetic load crosses the 50\,\% and 100\,\% thresholds smoothly, so every reclaim and feed is a clean step in the targets, and the three floors produce direct and indirect conflicts every cycle. On the testbed (Fig.~\ref{fig:simhw}(b)) the same reclaim-and-feed pattern appears over the same number of cycles, but utilisation jumps by tens of percentage points between cycles; this is why the three-cycle smoothing exists and why the pool fills and empties faster than in simulation. The simulation therefore validates the control logic, and the testbed validates it under real load; the two are not expected to agree numerically.

\subsection{Testbed: Spectral-Efficiency Behaviour}
\begin{figure}[t]
\centering
\includegraphics[width=0.9\columnwidth]{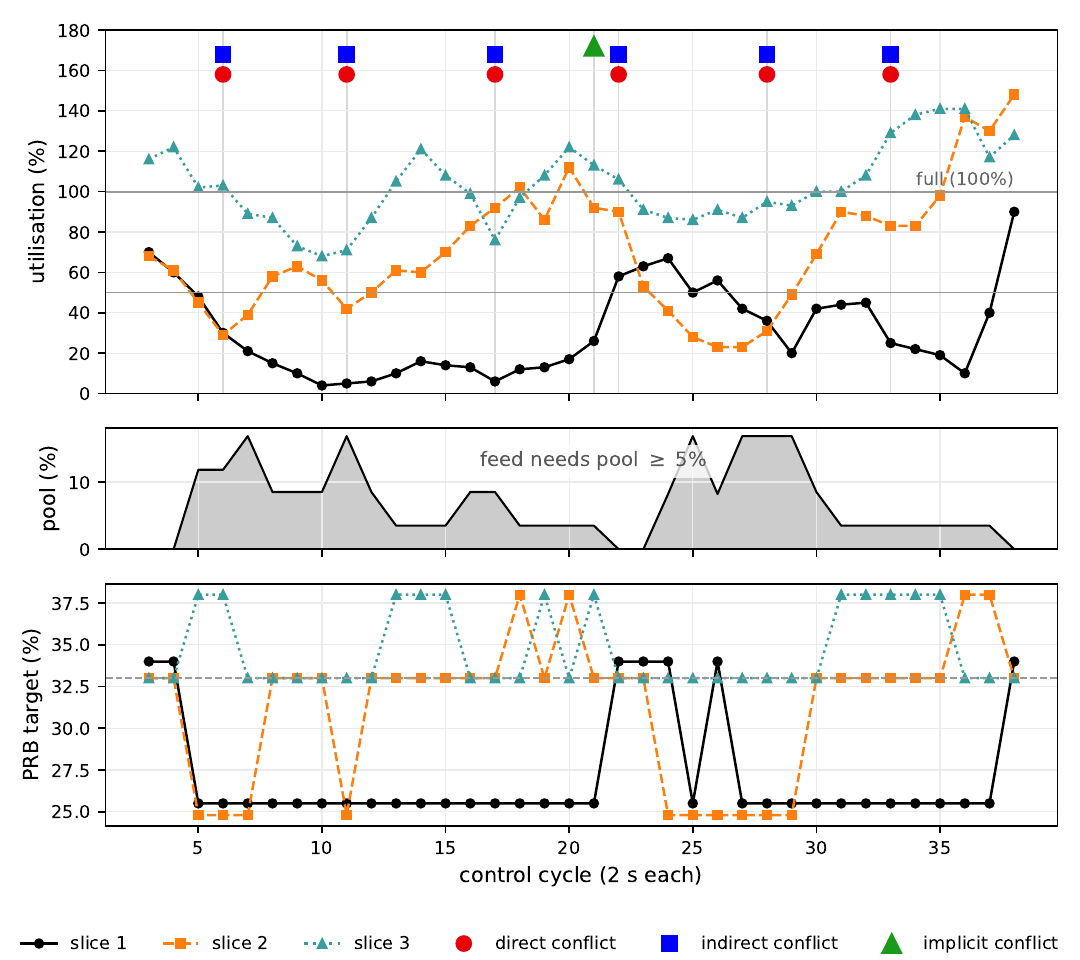}
\caption{Spectral-efficiency xApp on the testbed over a 36-cycle (134\,s) window. Top: per-slice utilisation with conflicts at their time of occurrence. Middle: the shared pool. Bottom: PRB targets.}

\label{fig:results}
\end{figure}

Fig.~\ref{fig:results} shows the spectral-efficiency xApp reallocating PRB capacity through the shared pool: under-utilised slices are reduced toward $0.75B_i$, and saturated slices receive five percentage points only when the pool holds at least five.

\begin{figure}[t]
\centering
\includegraphics[width=0.9\columnwidth]{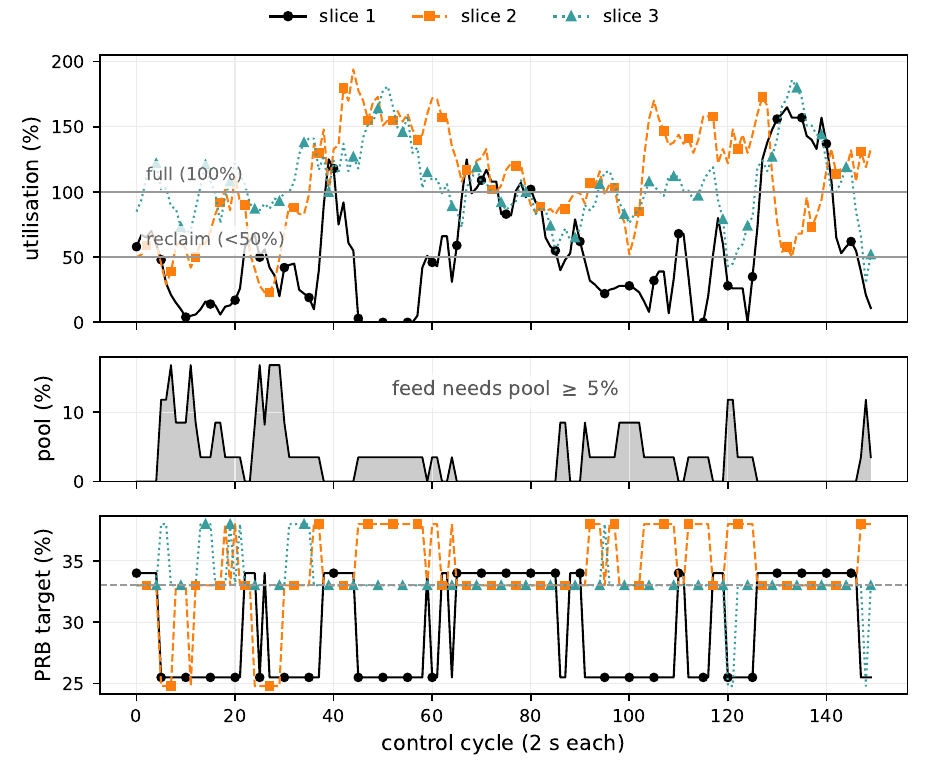}
\caption{Sustained operation over the first 150 cycles (561\,s) of the continuous run: utilisation (top), pool (middle) and PRB target (bottom).}
\label{fig:sepool}
\end{figure}

In one control cycle, slice~1 reached 116\,\% utilisation
relative to its baseline and was increased from 34\,\% to 39\,\%, while
slice~2, at 34\,\% utilisation, was reduced from 33\,\% to 24.8\,\%.
Slice~3 remained within the nominal utilisation range and retained its
33\,\% allocation. Table~\ref{tab:quant} compares the 36-cycle window of Fig.~\ref{fig:results} with the first 150 cycles of the continuous run in Fig.~\ref{fig:sepool}. The two are different load regimes rather than the same scenario at two time scales: the median utilisation rises from 68.5\,\% to 93.0\,\%, the share of slice-cycles in the saturated state doubles from 23.1\,\% to 44.9\,\%, and the share in the under-utilised state falls from 35.2\,\% to 20.7\,\%. Because the pool is filled only by under-utilised slices, the policy's reach collapses with load: a feed was possible (pool $\ge5$ points) in 47.2\,\% of cycles in the short window but in only 18.7\,\% of cycles in the continuous run. The feed rate is therefore bounded by the pool rather than by demand, and highly loaded slices can remain above their baseline utilisation after reallocation. The policy redistributes available capacity without eliminating overload.

Table~\ref{tab:quant} also compares the dynamic policy with the conventional static split of xApp~A (34/33/33\,\%, no reallocation), obtained by replaying the static allocation on the load trace measured under the dynamic policy; this is a counterfactual on the same trace, not a second live run. Relative to the static split, the dynamic policy reduces unmet demand by 33\,\% in the short window and by 12.5\,\% in the continuous run, and idle allocation by 23\,\% and 21\,\%, whereas the demand served rises by only about 2\,\% because the five-point feed is deliberately small. That the decisions reach the radio is confirmed over the full 276-cycle run: a $+5$-point target change was followed by a mean change in delivered usage of $+7.0\,\%$ in the next cycle ($n=78$), a $-8.25$-point change by $-6.4\,\%$ ($n=78$), and an unchanged target by $-0.02\,\%$ ($n=669$). Delivered usage nevertheless exceeded its own cap in 13\,\% (Fig.~\ref{fig:results}) and 39\,\% (Fig.~\ref{fig:sepool}) of slice-cycles, so the configured \texttt{max} share acts as a weighted share rather than a hard ceiling in this gNB build.

\begin{table}[t]
\caption{Quantitative comparison of the short window (Fig.~\ref{fig:results}) and the continuous run (Fig.~\ref{fig:sepool}), and of the dynamic policy against the static 34/33/33\,\% split replayed on the same load trace.}
\label{tab:quant}
\centering
\footnotesize
\setlength{\tabcolsep}{3pt}
\renewcommand{\arraystretch}{1.05}
\begin{tabular}{@{}lcc@{}}
\toprule
 & Fig.~\ref{fig:results} & Fig.~\ref{fig:sepool}\\
\midrule
Control cycles (measured duration) & 36 (134\,s) & 150 (561\,s)\\
Median / 90th-pct.\ utilisation $\rho_i$ & 68.5 / 121.0\,\% & 93.0 / 154.0\,\%\\
Slice-cycles saturated ($\rho_i\ge1$) & 23.1\,\% & 44.9\,\%\\
Slice-cycles under-utilised ($\rho_i<0.5$) & 35.2\,\% & 20.7\,\%\\
Mean pool & 6.9\,\% & 3.3\,\%\\
Cycles with pool $\ge5$ pts (feed possible) & 47.2\,\% & 18.7\,\%\\
\midrule
\multicolumn{3}{@{}l}{\emph{Per cycle, dynamic / static (\% of cell budget)}}\\
Demand served & 65.1 / 63.7 & 77.5 / 75.8\\
Unmet demand & 2.8 / 4.2 & 12.0 / 13.7\\
Idle allocation & 28.0 / 36.3 & 19.2 / 24.2\\
\bottomrule
\end{tabular}
\vspace{-6pt}
\end{table}

\subsection{Testbed: Conflicts}
\begin{figure}[t]
\centering
\includegraphics[width=0.85\columnwidth]{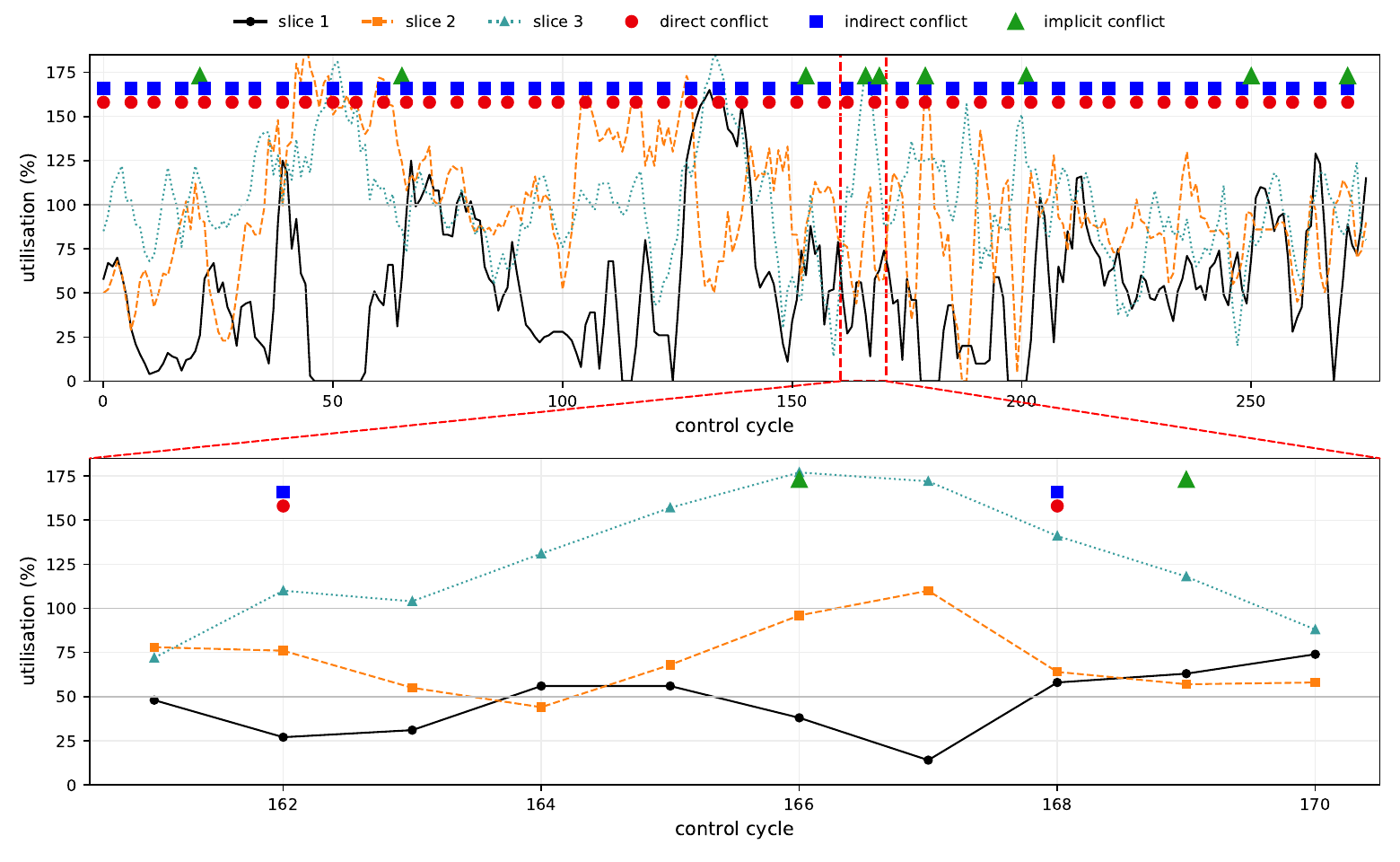}
\caption{Conflicts over the continuous run. Top: the full run with all conflicts
marked. Bottom: a 33\,s zoom of the boxed region where individual conflicts are
legible.}

\label{fig:zoom}
\end{figure}

We activated one instance of xApp~C per slice, each assigning a 40\,\% PRB floor to one UE, while xApp~B continued to update the slice-level caps. Under the operational definitions in Table~\ref{tab:conflicts_def}, the detector records a Direct event when $g_u>c_{s(u)}$, an Indirect event when $\sum_u g_u>100\,\%$, and an Implicit event when the slice target satisfies the defined oscillation criterion. Over the 276-cycle (17.1\,min) continuous run the scheduler fired 69 times, once every 15\,s. The detector de-duplicates an identical conflict signature within a 20\,s cooldown, so it logged 147 Direct, 49 Indirect, and 11 Implicit episodes (Table~\ref{tab:conflicts_def}); the Direct and Indirect counts are therefore bounded by run length divided by cooldown rather than by how often the condition held, which was every activation. As shown in Fig.~\ref{fig:zoom}, Direct and Indirect markers arrive in lock-step because each synchronous activation produces three per-slice floor--cap violations and one shared-budget violation, giving the $147=3\times49$ relationship, whereas the Implicit markers cluster where a utilisation trace crosses a threshold repeatedly, which is the visual reason for the caution applied to them in Section~\ref{sec:results}-D. The counts classify logged control intents and do not quantify UE-level impact; conflict resolution was disabled throughout.

\subsection{Budget Conservation and Robustness}
Three checks address whether the budget and correct UE-level control hold under delayed, rejected, or identifier-perturbed updates. First, conservation~\eqref{eq:cons} holds exactly in all 276 cycles of the policy arithmetic, and also after the integer quantisation applied at the actuator: the pushed slice shares summed to 100 in 127 cycles and to 97, 92, 89, or 84 in the remaining 149, never above 100, the shortfall being the withheld pool. Second, all 1035 commands (828 slice, 207 UE) reached the gNB with zero transport failures or timeouts; each push is a synchronous request--response, and the gap between consecutive commands within a cycle, which bounds the round trip, had a 95th percentile of 1\,ms ($n=552$, the timestamp resolution), so delayed updates are not a factor here. Third, the gNB rejected 11 of the 207 UE pushes (5.3\,\%) in five episodes, because xApp~C addresses a UE by a static gNB-DU-UE-F1AP index that goes stale when the handset re-attaches; every episode recovered on the next activation, and 59 of the 69 activations had all three floors accepted. No slice push was rejected, since slices are addressed by SST, and the core-side load proxy was unaffected because it is keyed by subscriber and its address-to-slice map is rebuilt every 10\,s. UE-level control is thus identifier-fragile on this platform whereas slice-level control and the measurement are not; the detector evaluates logged intents, so the rejected pushes are included in the Direct counts.

\subsection{Discussion}

The spectral-efficiency control performs the intended reclaim-and-feed operation under real traffic, although hardware behaviour is far more variable than in simulation: the testbed exhibits rapid fluctuations and strong load asymmetry across slices, and Table~\ref{tab:quant} quantifies how this limits the policy's reach.

Utilisation above $100\,\%$ means the usage proxy exceeds the fixed baseline ($\rho_i=u_i/B_i$), not physical PRB use beyond the radio resources. Shared-budget violations arise only when the three $40\,\%$ UE guarantees are considered jointly, since xApp~B itself conserves the budget.

The 11 events labelled Implicit should be interpreted more cautiously. They are allocation oscillations satisfying the detector's operational criterion, but the experiments do not establish that they are caused by inter-xApp interaction, since threshold crossings driven by variable traffic can produce the same behaviour. We therefore report them as Implicit under our operational interpretation rather than as definitive evidence of an O-RAN implicit conflict.
Finally, the conflict counts quantify incompatibilities among logged control intents, not their end-user consequences; UE-level throughput, latency, or service degradation per event is not measured here.


\section{Conclusion}\label{sec:conclusion}

This work experimentally investigated cross-granularity xApp interaction between slice-level PRB control and per-UE scheduling in O-RAN. The spectral-efficiency xApp reclaimed capacity from under-utilised slices and redistributed it to saturated slices while conserving the cell-wide PRB budget in every cycle, before and after actuator quantisation. Against the static slice split replayed on the same load trace, it reduced unmet demand by 12--33\,\% and idle allocation by 21--23\,\%, with its reach bounded by the pool under heavy load. On the hardware testbed, the detector consistently identified per-slice floor--cap violations and shared-budget over-subscription (147 and 49 events, the expected $3{:}1$ ratio). Allocation oscillations were observed under variable traffic but not attributed to inter-xApp interaction, and UE-level control proved identifier-fragile where slice-level control was not. The reported conflicts are incompatibilities among control intents rather than measured UE-level degradation; future work will measure UE-level impact, isolate the causal interaction between competing xApps, and evaluate mitigation strategies on the testbed. As the RAN moves towards AI-native control, the logged intents and detector labels produced here are the hardware-derived training data that learned conflict detection~\cite{wadud2026ai} will require.
\section*{Acknowledgment}
This work was supported by Research Ireland CONNECT-2 grant 13/RC/2077-P2.
\bibliographystyle{IEEEtran}
\bibliography{references}

\end{document}